\documentclass[11pt,twoside]{article}
\usepackage{asp2010}
\usepackage{hyperref}
\usepackage{bmpsize}

\resetcounters
\begin{document}

\title{ADS 2.0: new architecture, API and services}
\author{
Roman Chyla,
Alberto Accomazzi, 
Alexandra Holachek, 
Carolyn S. Grant,
Jonathan Elliott,
Edwin A. Henneken, 
Donna M. Thompson, 
Michael J. Kurtz, 
Stephen S. Murray,
Vladimir Sudilovsky
\affil{Harvard-Smithsonian Center for Astrophysics, 60 Garden Street, Cambridge, MA 02138, USA}}

\begin{abstract}
The ADS platform is undergoing the biggest rewrite of its 20-year history.  While several components have been added to its architecture over the past couple of years, this talk will concentrate on the underpinnings of ADS's search layer and its API. To illustrate the design of the components in the new system, we will show how the new ADS user interface is built exclusively on top of the API using RESTful web services. Taking one step further, we will discuss how we plan to expose the treasure trove of information hosted by ADS (10 million records and fulltext for much of the Astronomy and Physics refereed literature) to partners interested in using this API.  This will provide you (and your intelligent applications) with access to ADS's underlying data to enable the extraction of new knowledge and the ingestion of these results back into the ADS. Using this framework, researchers could run controlled experiments with content extraction, machine learning, natural language processing, etc. In this talk, we will discuss what is already implemented, what will be available soon, and where we are going next.
\end{abstract}

\section{Introduction}

The SAO/NASA Astrophysics Data System \citep{2000A&AS..143...41K} has gone through significant changes over the course of its 20 year career, but its purpose has always been to provide relevant information and services to the scientific community. The ADS has achieved this goal through constant innovation and adoption of new technologies. In this paper we will describe the changes that are happening to the ADS in its recent incarnation. They constitute the biggest rewrite in the history of the system and mark the opening of many new possibilities, particularly in discovery capabilities and support for bibliometric studies.

\section{Environment in which the ADS operates}

The ADS is an \emph{information system} - built by and for scientists to provide access to scientific literature. Its user base is considerable, currently at any given time there are $\sim200-300$ active users on the website (which amounts to $\sim1$ million unique sessions per month), although its core user base is much smaller \citep{2009astro2010P..28K}. For the majority of these (anonymous) users worldwide the ADS is the destination they reached after following a link from Wikipedia or Google to one of its records. However for about 50,000 scientists worldwide, the ADS represents a gateway to information, they use the system every day (or every other day). And for the ADS they are the users who really care and need to be cared for.

The ADS provides its services within a greater information space environment, which includes several service providers. One such providers is the arXiv,  arguably one of the most indispensable academic services, at least in the fields of astronomy and physics. It has become a de facto publishing standard to post pre-prints on the arXiv before they are available as published and reviewed papers through more traditional channels. However, as many things that we take for granted and expect them to be there, the arXiv operates with a very limited budget and resources.

The arXiv (and its volunteers, the scientists) performs the hard work of pre-filtering submissions and categorizing them. It does serve one purpose and serves it well, yet it needs other projects to complement it. One such project is the ADS, which closely collaborates with arXiv, downloads and indexes every single arXiv submission - it is perhaps a little known fact, but you can use the powerful search capabilities of the ADS to sift through all arXiv fulltexts.

Throughout its existence, the ADS has established such symbiotic relationships in order to provide the best service to the scientific community. It collaborates not only with arXiv, but also with data repositories, publishers, and people behind production of journals and conferences, it has linked pre-prints to the published papers and other grey-literature, so that they not disappear into obscurity. Tracking the important channels of scientific information logically led to demands for indicators of scientific contribution (the citation indicators) - which in turn get (mis)used as metrics of scientific success. Despite how we may look at it, these indicators remain important tools and the ADS recognizes their importance. In addition to scientists, who come in their roles of researchers and readers, there are also administrators and evaluators, librarians, as well as technocrats and grant agencies who use the ADS database for other, non-scientific purposes. \citep{2005JASIS..56..111K,2007ASPC..377...69A}

The reason we mention these diffeent user populations is to explain the context in which the ADS operates. It has to choose wisely which services are feasible to implement and maintain. Given limited resources, this becomes not only a technological, but also a sociological challenge.

The discussion becomes simpler if we consider the entities (objects) processed by the ADS. As an information system, the ADS primarily processes \emph{scientific literature}, so the most frequent data types that it deals with are: 

\begin{itemize}
\item \emph{Preprints} and \emph{journal articles}: The ADS will store them in the database (their \emph{fulltext} content as well as attached objects such as \emph{extracted images}).
\item \emph{Metadata}: Metadata can be very expensive to create and curate but its use may be surprisingly powerful. For example, if author affiliations are known, one can compile cumulative reports about the scientific output of a certain institute, department and/or university (even country or continent).
\item \emph{Citation information}: Citing and cited by are very powerful pieces of information, they are the basis of the citation index and the metrics available throughout the ADS. Its use in the ADS is so important that in the new version of the search engine it has been incorporated into the index, so you can search for citations in the same way as you search for papers.
\item \emph{Usage information}: This is information about \emph{readers}, what they are reading, what papers are popular, downloaded, consulted and how often.\footnote{This information is  properly anonymized before it gets used by different algorithms and subsystems of the ADS.}
\end{itemize}

Finally, there is \emph{raw data}, but in case of the ADS rather \emph{Links to science data}. These consists of a network of links between bibliographic records hosted by ADS and data products, astronomical object, and electronic articles hosted elsewhere. Other projects (such as NED and the CDS) specialize in collecting this information and do a much better job with it than what the ADS could do. The role of the ADS here is to serve as a gateway and make these resources \textbf{seen and used}.

And what we mean by saying that is not just people who sit behind computer screens looking at the ADS web interface - to the ADS developers those people are morphing with the machines (robots) that start to access the new ADS infrastructure. This has nothing to do with de-humanization or other negative connotations of information processing. As far as information systems are concerned, it has become impractical to treat users differently from machines. The differences between the humans and the clients (robots, programs, machines) that access the ADS on the users' behalf are disappearing, and this brings us to the new era and \textbf{new architecture} of the ADS. Since a few years, the system has been preparing itself for a change and it is not exactly \emph{evolution} but rather \emph{revolution}.

\section{New architecture and the API}

While the old infrastructure of the ADS is similar to a huge building which hosts every department and which has only one entrance (and only one power supply), the new infrastructure is much more distributed and lightweight, more similar to a town with roads connecting different buildings. The central component here is the \emph{Application Programming Interface (API) layer} \footnote{The API layer is itself published on GitHub, at \url{http://github.com/adsabs/adsws}} - we have split the internal subsystems into \emph{microservices}. They are independent, standalone web-services that communicate with each other through the API using the REST and OAuth protocols (and soon also exclusively through HTTPS). Every service, even the ones that we consider the most critical, are connected to the API and available through it; can be accessed from anywhere on the internet provided that the client has the proper authorization token.

The micro-services are virtualized, currently running off the Amazon cloud. They  can be scaled horizontally: if there is need for more power, we add computational nodes to the cloud. Using this new organization, the ADS can more easily add new functionality (new services), and it is relatively easy to alter existing services. Basically, implementations can be swapped at will, provided the semantics of the API remains unchanged. It is still challenging to find the absolute minimal set of the API, but the system has already proven itself. It simplifies maintenance and allows for rapid development. \footnote{As will be discussed later, this has also a negative consequences; at the moment, the ADS is changing so rapidly, that is hard to keep the documentation up to date. It requires certain mental discipline and robust practices and we are still struggling to find the optimum balance.}

\subsection{The search}

The core of the new ADS (similarly to the previous generation of the ADS) is centered around \textbf{search}. The search engine is built on a heavily modified and extended version of Apache SOLR (\url{http://lucene.apache.org/solr/}). As one would probably expect, it has many advantages over the in-house custom search service  developed for ADS Classic 20 years ago. Unsurprisingly, it is able to address much bigger memory spaces which translates into an ability to index bigger volumes of data. For the first time in ADS history, it is possible to search through fulltext content together with the metadata. Additionally, we have integrated the citation and co-readership networks into the index, so you can now simultaneously explore all three different search spaces: metadata, fulltext content, and citations.

The new search engine provides many features, some of which will be totally new to the ADS users, for example:

\begin{itemize}
\item proximity search -- ability to find a group of tokens based on their distance, e.g. body:(weak NEAR5 lensing), which finds phrases where “weak” appear within 5 words of “lensing.” This query would find phrase "weak lensing", as well as "weak gravitational lensing" in the fulltext
\item similarity search -- (e.g. author:eisenstein$\sim0.3$) will find different spellings of the author name based on different similarity metric (this can be used to correct for misspellings)
\item regular expressions -- most often used by curators to find terms that follow a particular pattern e.g. grant numbers or astronomical entities
\end{itemize}

The features described above are provided by the Lucene search library upon which SOLR is built.  In addition, ADS has extended Lucene capabilities by implementing its own \emph{query language}, modifying the traditional search syntax and  enhancing it with new search operators and modifiers.  The main purpose of this “dialect” is to facilitate communication between clients and servers.  We do not expect normal ADS users to know the specialized syntax, but we want to give them tools to execute complex queries if they ever have a need to do so. More likely though, this query language will be used by clients (programs) that search the ADS on behalf of users. To illustrate the point, here is a short list of a few specialized modifiers/operators:

\begin{itemize}
\item =searchterm -- the equals sign modifier is perhaps known to ADS users, it disables the synonym expansion so that the search engine will not look for synonyms or alternate spellings.
\item references(author:einstein) and citations(author:einstein) -- these are the operators to retrieve papers that cite a collection of papers, or those that are cited by that collection. Operators can be freely nested and combined with other terms, even if that means that you are going to analyze the entire ADS corpus, e.g. citations(references(author:einstein))
\item useful(), instructive(), trending() and other operators -- the ADS search engine supports so-called second order queries which use the citation and readership network to provide advanced search capabilities.  Currently these operators use multiple algorithms to retrieve papers that are most cited by papers on a topic (useful), citing the most relevant papers on a topic (instructive), or most read by people interested on a topic (trending).  The ADS will be experimenting with different implementations of these operators and will be adding more of them as needed.
\end{itemize}

The citation operators together with custom analytical functions are potentially the most fun part of the new search engine. It is possible to construct custom searches just for the purpose of supporting certain visualizations (e.g. a view of a co-readership network) or use cases (e.g. selecting the \emph{N}th token in an list of author names). In the future, we expect users of the API to come up with novel ways of using them, something we did not anticipate when they were created. \footnote{Of course we are aware that in order for this to happen we need to provide better documentation of the search functionality.}

\subsection{Other services}

The best example of the API in action is the new user interface for the ADS, code-named Bumblebee. It is a rich JavaScript client that runs in your web browser. It is the future replacement of the ADS Abstract service, but unlike that one, virtually every interaction happens through the API. So without knowing it, when you visit \url{http://ui.adslabs.org}, you are automatically becoming an API user and accessing multiple web services. It will not feel any different, but the difference in implementation is significant: to have access to the API means that you are a first-class citizen of the ADS world, accessing the same interface that ADS developers use every day.

\begin{figure}
\centering
\includegraphics[width=1\textwidth,natwidth=1178,natheight=510]{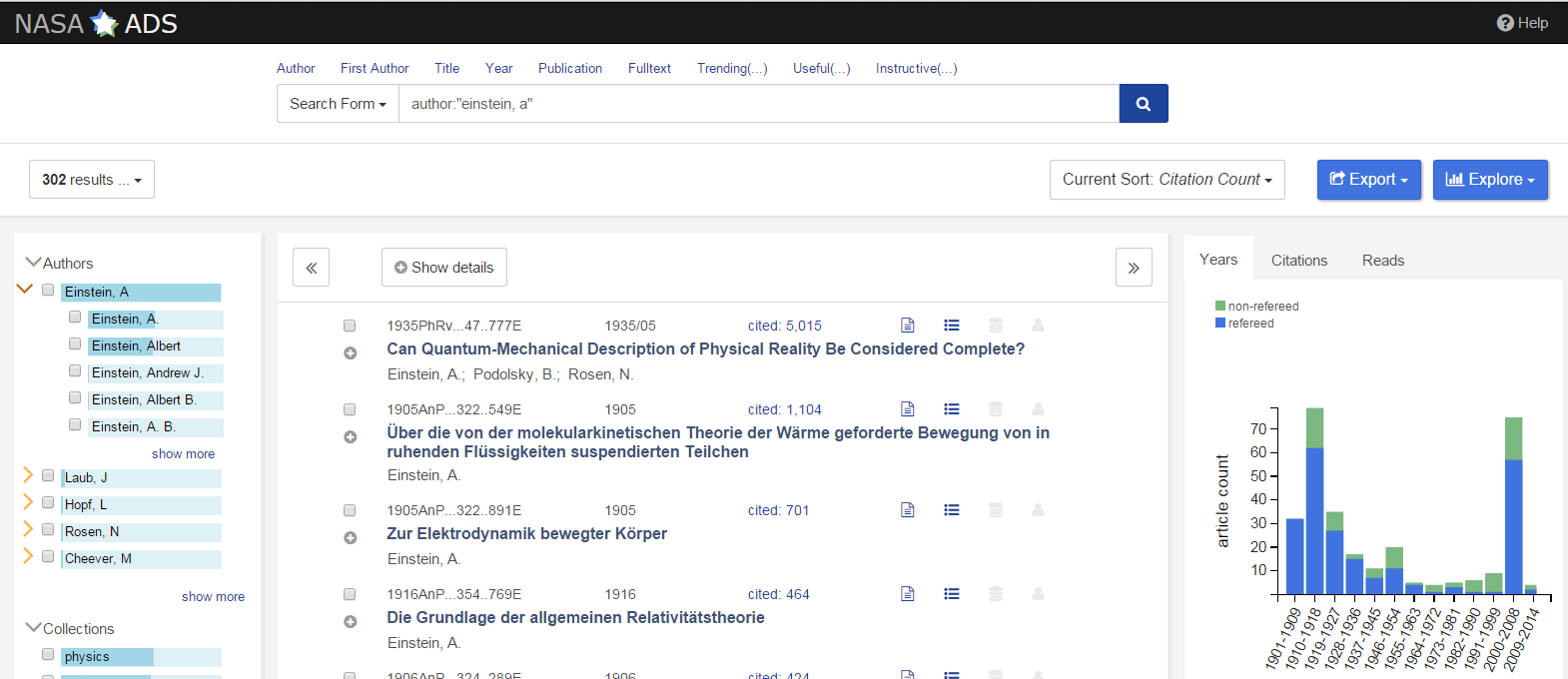}
\caption{\label{fig:bumblebee}Screenshot of the new UI for the ADS, using normal search.}
\end{figure}

\begin{figure}
\centering
\includegraphics[width=1\textwidth,natwidth=982,natheight=652]{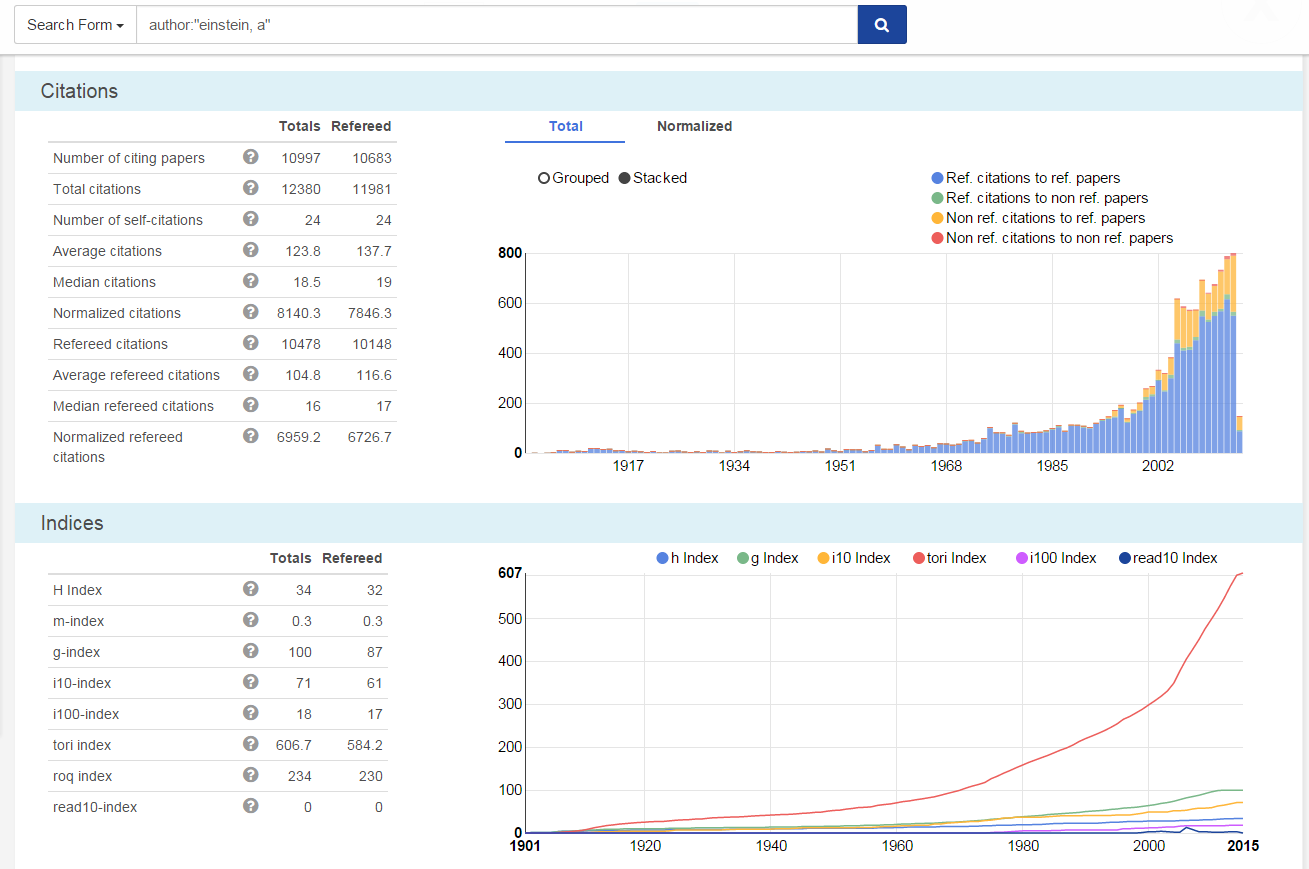}
\caption{\label{fig:metrics}Citation metrics report for the most cited 100 papers of A.~Einstein.}
\end{figure}

\begin{figure}
\centering
\includegraphics[width=1\textwidth,natwidth=976,natheight=630]{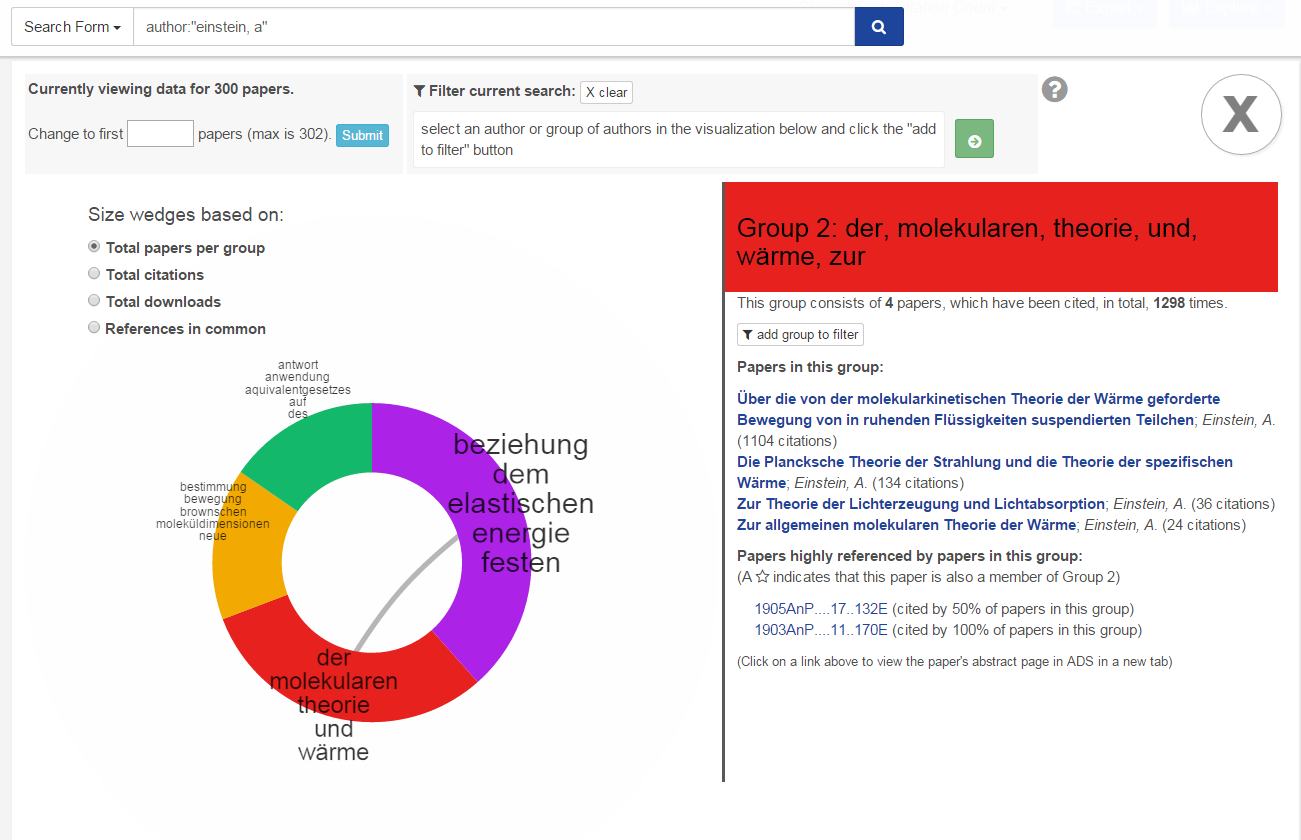}
\caption{\label{fig:paper-network}Paper networks inside the first 100 most cited papers of A.~Einstein.}
\end{figure}

At the time of this writing, there exist multiple services besides search:

\begin{itemize}
\item \emph{metrics}: the familiar citation and usage reports (see \ref{fig:metrics}), if you access the API directly, you can download the data that are used to generate the metrics views (and build your own version of the visualization)
\item \emph{visualizations}: this service exposes different visualizations based on the relations between authors and papers, e.g. word-clouds, paper networks, citation networks, and bubble-charts of search results (see \ref{fig:paper-network}).
\item \emph{ORCID}: service that allows users to claim authorship through the ADS interface into the ORCID database\footnote{ORCID stands for Open Researcher and Contributor ID: \url{http://orcid.org}}
\end{itemize}

Information on how to access and use the Api for these services, and any future ones, is available through: \url{http://adsabs.github.io/help}.

\section{Future directions}

We are aware that the new API has a number of shortcomings especially with respect to documentation. It is still a very young and relatively new component and it may not be a smooth ride for external developers, but the ADS has already gone a long way in making this new "mode" available and will continue improving it. In the past year, the ADS has re-engineered many of its subsystems from the ground up (after a few false starts).  The current implementation has shown that the system architecture is robust, and this gives us confidence to keep going in this direction. To offer a level of stability to the platform, major changes to the API will be versioned so that we will be able to provide backward compatibility to applications built on it.  Our hope is that ultimately this open API platform will attract development from outside the ADS group, leveraging potential hidden in wider community.

The new generations of astronomers and data scientists are technically very savvy and ready to embrace emerging technologies to conduct their research.  We hope that they will appreciate the level of access the new API provides, and that they will take the opportunity to build something that satisfies their research needs (and share it with other people). We hope that somebody will come up with ideas we did not have the time or acuity to consider, new ways to explore the vast data indexed in ADS, or new ways to connect existing systems and give insights that were not previously possible. We will continue building and adding more services, in the near future improving the relevancy of the search, and preparing the new user interface to replace the existing ADS Abstract Service. Once the new user interface is rolled out, any new contributions to it will become visible to at least a million users every month, which will hopefully provide enough usability data to test even more of the new ideas. Intrigued and want see more or perhaps even build something?  All of the code is published open-source and available through ADS’s Github repo: \url{http://github.com/adsabs}.  We welcome any and all comments, criticisms and contributions.

\acknowledgements This work has been supported by the NASA Astrophysics Data System project, funded by NASA grant NNX12AG54G.

\bibliography{07_1}

\end{document}